\documentclass[%
 reprint,
 amsmath,amssymb,
 aps,
 prc,
]{revtex4-2}

\usepackage{graphicx}
\usepackage{dcolumn}
\usepackage{bm}
\usepackage{rotating}
\usepackage{amsmath}
\usepackage[hidelinks]{hyperref}
\usepackage[dvipsnames]{xcolor}

\newcommand{\be}{\begin{equation}}
\newcommand{\ee}{\end{equation}} 
\newcommand{\bea}{\begin{eqnarray}}
\newcommand{\eea}{\end{eqnarray}} 
\newcommand{\bitz}{\begin{itemize}}
\newcommand{\eitz}{\end{itemize}}
\newcommand{\ben}{\begin{enumerate}}
\newcommand{\een}{\end{enumerate}} 
\newcommand{\bdesc}{\begin{description}}
\newcommand{\edesc}{\end{description}} 
\newcommand{\idest}{{\it i.e.\ }}

\newcommand{\nn}{\nonumber} 
\newcommand{\jpi}{$J\pi$ }

\newcommand{\we}{Weisskopf-Ewing }

\newcommand{\cn}{compound nucleus }

\begin{document} 

\title{Cross sections for neutron-induced reactions from surrogate data:
revisiting the Weisskopf-Ewing approximation for (n,n') and (n,2n)
reactions}

\author{Oliver C. Gorton}
\email{ogorton@sdsu.edu} 
\affiliation{ San Diego State University, San Diego, California 92182, USA }
\altaffiliation{ University of California, Irvine, California 92679, USA }
\author{Jutta E. Escher}
\email{escher1@llnl.gov}
\affiliation{ Lawrence Livermore National Laboratory, Livermore, CA 94550}

\date{\today}

\begin{abstract} 
\begin{description}
\item[Background] Modeling nuclear reaction
networks for nuclear science applications and for simulations of astrophysical
environments relies on cross section data for a vast number of reactions, many
of which have never been measured. Cross sections for neutron-induced reactions
on unstable nuclei are particularly scarce, since they are the most difficult to
measure. Consequently, we must rely on theoretical predictions or indirect
measurements to obtain the requisite reaction data. For compound nuclear
reactions, the surrogate reaction method can be used to determine many cross
sections of interest.
\item[Purpose] Earlier work has demonstrated that cross sections for
neutron-induced fission and radiative neutron capture can be determined from a
combination of surrogate reaction data and theory. For the fission case, it was
shown that Weisskopf-Ewing approximation, which significantly simplifies the
implementation of the surrogate method, can be employed. Capture cross
sections cannot be obtained, and require a detailed description of the
surrogate reaction process. In this paper we examine the validity of the
Weisskopf-Ewing approximation for determining unknown $(n,n^{\prime})$ and
$(n,2n)$ cross sections from surrogate data.
\item[Methods] Using statistical reaction calculations with
realistic parametrizations, we investigate first whether the assumptions
underlying the Weisskopf-Ewing approximation are valid for $(n,n^{\prime})$ and
$(n,2n)$ reactions on representative target nuclei. We then produce simulated
surrogate reaction data and assess the impact of applying the Weisskopf-Ewing
approximation when extracting $(n,n^{\prime})$ and $(n,2n)$ cross sections in
situations where the approximation is not strictly justified.
\item[Results] We find that peak cross sections can be estimated
using the Weisskopf-Ewing approximation, but the shape of the $(n,n^{\prime})$
and $(n,2n)$ cross sections, especially for low neutron energies, cannot be
reliably determined without accounting for the angular-momentum differences
between the neutron-induced and surrogate reaction.
\item[Conclusions] To obtain reliable $(n,n^{\prime})$ and $(n,2n)$
cross sections from surrogate reaction data, a detailed description of the
surrogate reaction mechanisms is required. To do so for the compound-nucleus
energies and decay channels relevant to these reactions, it becomes necessary to
extend current modeling capabilities. 
\end{description}
\end{abstract}

\maketitle

\section{Background and Need} \label{sec_intro}

Nuclear reaction data are required for many applications in both basic and
applied science, whether it be for modeling the origin of elements in the
universe, the safe operation of a next-generation reactors, or for
national-security applications \cite{Arcones:17,Hayes:17}. Nuclear reaction
libraries provide evaluated reaction data for many such applications
\cite{Capote:09}. These evaluations are based on nuclear reaction calculations
anchored to experimental data and state-of-the-art nuclear theory. As many
reaction cross sections of interest cannot be measured directly, due to short
lifetimes or high radioactivity of the target nuclei involved, indirect methods
are being developed~\cite{Baur:96,Typel:03,Escher:12rmp,Larsen:19} to address
the gaps and shortcomings in present databases.

In this paper we focus on the ``surrogate reaction method''
\cite{Escher:12rmp,Escher:16a}, an indirect approach for determining cross
sections for compound-nuclear reactions. Compound-nuclear, or ``statistical''
reactions, proceed through the formation of an intermediate ``compound''
nucleus $n+A \to B^*$, followed by a decay into reaction products $B^*$
$\rightarrow$ $c + C$. The appropriate formalism for calculating cross sections
for these reactions is the Hauser-Feshbach formalism~\cite{HauserFeshbach:52,
Froebrich:96}. Hauser-Feshbach calculations are often quite limited in accuracy
due to uncertainties in the nuclear physics inputs needed, in particular the
nuclear structure inputs associated with the decay of the compound nucleus (CN).

\begin{figure}[htb] \begin{center}
\includegraphics[width=\columnwidth]{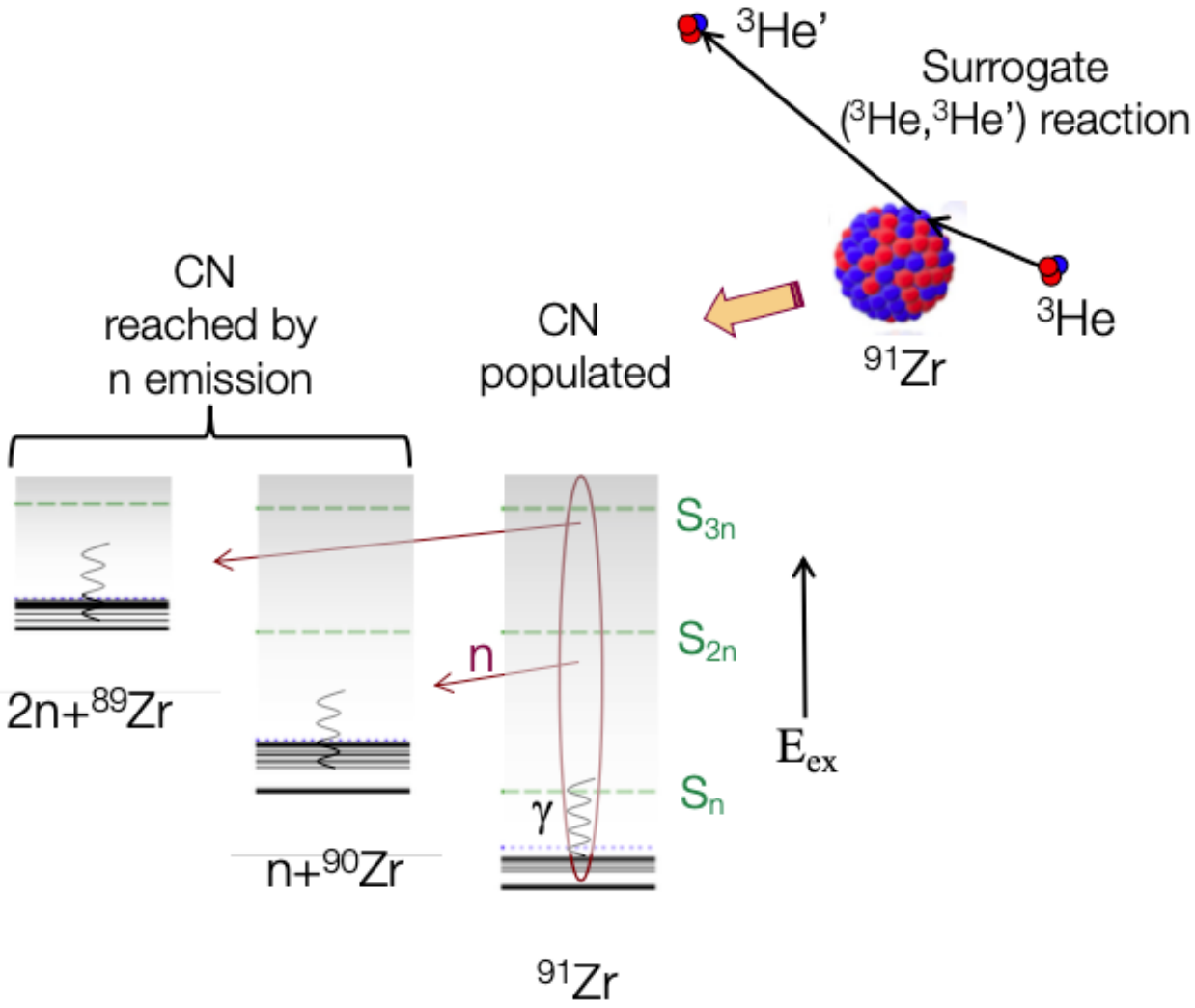} 
\end{center} 
\caption{Surrogate reactions approach for the simultaneous measurement of
$^{90}$Zr$(n,\gamma)$, $^{90}$Zr$(n,n^{\prime})$, and $^{90}$Zr$(n,2n)$ cross
sections. A recent inelastic scattering experiment produced the CN up to
about 30 MeV, \idest above the two-neutron threshold \cite{Scielzo:priv}.
Subsequent decay via emission of $\gamma$s, one neutron, and two neutrons,
produces final $^{91}$Zr, $^{90}$Zr, and $^{89}$Zr nuclei, respectively. The
example here displays a situation in which discrete $\gamma$ transitions between
low-lying states in three nuclei are used to determine the decay channel
probabilities. A complementary decay measurement that focuses on the detection
of neutrons is under development as well~\cite{Hughes:00ip}. The $^{90}$Zr
experiment serves as a benchmark, since multiple neutron-induced reactions for
the stable $^{90}$Zr nucleus are known from direct
measurements~\cite{Semkova:10}.}
\label{Fig_n2nSurrogate} 
\end{figure}

In a surrogate reaction experiment, the CN of interest is produced via an
alternative, experimentally accessible reaction, and the probability of decay
into the reaction channel of interest is measured. From this data, constraints
for the Hauser-Feshbach calculations can be obtained.

The surrogate method has some significant advantages over alternative indirect
approaches: 1) the method does not require measurement of auxiliary nuclear
properties that are not available for unstable nuclei and for which
interpolation or extrapolation procedures are associated with uncontrolled
uncertainties~\cite{Escher:18prl,Ratkiewicz:19prl}, and 2) The method can be
used for reactions that populate energies well above particle separation
thresholds in the CN, \idest it is applicable not only to $(n,\gamma)$, but
also to $(n,n^{\prime})$, $(n,2n)$, $(n,p)$, $(n,f)$ reactions (and similarly
to charged-particle-induced reactions).

Alternative indirect approaches, in particular the Oslo and $\beta$-Oslo methods
\cite{Larsen:19}, aim at extracting level densities and $\gamma$-ray strength
functions by populating a CN below the neutron separation energy via a transfer
reaction or $\beta$ decay, respectively, and measuring the resulting gamma
emission. To separate the level density from the gamma-ray strength function,
the Oslo-type analyses require the use of additional information; typically,
this includes average neutron resonance spacings ($D_0$) and the average
radiative widths, $\langle \Gamma_\gamma \rangle$. For neutron-induced
reactions on unstable nuclei, however, these quantities are not available and
are difficult to estimate reliably. In addition, the (n,n’) and (n,2n)
reactions of interest here require CN decay information for excitation energies
well above the neutron separation energy.

Both the surrogate method and the Oslo/$\beta$-Oslo methods require the
calculation of the formation of the CN in the desired reaction. This involves,
for neutron-induced reactions, knowledge of a neutron-nucleus optical model
potential. For target nuclei near stability, global nucleon-nucleus optical
models exist \cite{Koning:03, Varner:1991}, which are expected to be reliable
at least a few isotopes away from stability. While these optical models are
often applied far from stability, little is known about how well they work in
these areas of the isotopic chart \cite{Johnson_2020, PhysRevC.107.014602,
Hebborn2023}. More theoretical work is needed to
develop the next generation of optical model potentials. These need to display
the proper dispersive properties and reflect the correct isospin dependence,
and are ideally based on microscopic theories \cite{Dickhoff:19, Pruitt:20b,
PhysRevLett.127.182502, Blanchon:2015}. In addition, new experiments at
radioactive beam facilities are needed to constrain and test the optical
models. 

Applications of the surrogate method to $(n,f)$ reactions have a long history
\cite{Escher:12rmp} and in recent years scientists successfully used the
approach to obtain neutron capture cross sections
~\cite{Escher:18prl,Ratkiewicz:19prl,PerezSanchez:20}. In this paper, we focus
on possible applications to $(n,n^{\prime})$ and $(n,2n)$ reactions.

Figure~\ref{Fig_n2nSurrogate} illustrates how the surrogate approach can be
used to determine $^{90}$Zr$(n,\gamma)$, $^{90}$Zr$(n,n^{\prime})$, and
$^{90}$Zr$(n,2n)$ cross sections from a surrogate inelastic scattering
experiment. For incident neutron energies below a few MeV, neutron capture and
inelastic neutron scattering compete with each other, above $E_n \approx $ 10
MeV, one- and two-neutron emission compete with each other. Proton and
$\alpha$ emission compete only weakly and have to be accounted for, but are not
shown here. In actinides, fission may compete at all energies. If the
surrogate reaction measurement is designed to cover a broad energy range, it
becomes possible to determine cross sections for all three neutron-induced
reactions in {\em one} experiment. The decay channel of interest is determined
either by measuring $\gamma$ transitions specific to one of the three decay
products, or by detecting outgoing neutrons, in coincidence with the
scattered $^{3}$He particle. Experimentalists conducting these measurements
have utilized discrete $\gamma$ rays and are currently developing the
capability to use neutron measurements.

In principle, a careful description of the surrogate reaction mechanism is
required to obtain the cross section of the desired reaction. This is because
one must account for the differences in the decay of the CN due to the
angular-momentum and parity differences in the surrogate and desired reactions
(the spin-parity mismatch). Indeed, $(n,\gamma)$ reactions are very sensitive
to spin effects, particularly in nuclei with low level density \cite{Chiba:10,
EscherDietrich:10, Forssen:07}. On the other hand, sensitivity studies 
for surrogate $(n,f)$ applications have shown that neglecting the spin-parity
mismatch yields reasonable results, except at low neutron
energies~\cite{Younes:03a,Younes:03b,EscherDietrich:06}. Neglecting the
spin-parity mismatch between the surrogate and desired reactions is known as the
Weisskopf-Ewing approximation, and it greatly simplifies the extraction of the
cross sections from surrogate data, as only a simple theoretical treatment is
required.

It is the purpose of this paper to investigate what is required to determine
reliable cross sections for $(n,n^{\prime})$ and $(n,2n)$ reactions from
surrogate data. Specifically, we carry out sensitivity studies that examine the
validity of the \we approximation for these two reactions for several regions of
the nuclear chart.

In the next section, we review the surrogate reaction formalism and provide
details on the Weisskopf-Ewing approximation. In Section~\ref{sec_method}, we
describe our procedure for testing the assumption of the approximation, and for
investigating the consequences of applying the approximation in situations where
its assumptions are not strictly valid. In Section~\ref{sec_results}, we present
results for zirconium, gadolinium and uranium nuclei, which are representative
of spherical and deformed nuclei, respectively. We summarize our findings and
make recommendations in Section~\ref{sec_outlook}.

\section{Reaction Formalism} \label{sec_formalism}

Here we summarize the Hauser-Feshbach formalism for calculating the cross
section of a compound-nuclear reaction and its relationship to the description
of a surrogate reaction. This clarifies how surrogate reaction data can be used
to constrain calculations for unknown cross sections. We outline the
circumstances under which the Weisskopf-Ewing approximation can be used to
simplify the analysis used to obtain the desired compound cross section.

\subsection{Theory for the desired reaction} \label{sec_formalism_desired}

The Hauser-Feshbach (HF) statistical reaction formalism properly accounts for
conservation of angular momentum and parity in compound-nuclear reactions. For
a reaction with entrance channel $\alpha = a+A$ that forms the CN $B^*$, which
subsequently decays into the exit channel $\chi = c+C$, 
\begin{equation*}
 a + A \to B^* \to c + C,
\end{equation*} 
the HF cross section can be written as
\begin{equation}\label{eq:DesReact}
 \sigma_{\alpha\chi}(E_a) = \sum_{J,\pi}
 \sigma_{\alpha}^{CN}(E_{ex},J^\pi)G_{\chi}^{CN}(E_{ex},J^\pi).
\end{equation}
Here $E_a$ and $E_{ex}$ are the kinetic energy of the projectile a and the
excitation energy of the compound nucleus $B^*$, respectively. They are related
to each other via $E_{a}=\frac{m_A}{m_a+m_A}(E_{ex}-S_{a})$, where $S_a$ is 
the energy needed to separate the particle $a$ from the nucleus $B^*$. $m_a$ 
and $m_A$ are the masses of the projectile and target, respectively. $J$ and 
$\pi$ are the spin and parity of the compound nucleus and
$\sigma_{\alpha}^{CN}(E_{ex},J^\pi)$ is the cross section for the forming the
\cn $B^*$ with spin and parity $J^\pi$ at energy $E_{ex}$. The
$\sigma_{\alpha}^{CN}(E_{ex},J^\pi)$ and their sum, the compound-formation cross
section $\sigma_{\alpha}^{CN} (E_{ex}) =$ $ \sum_{J,\pi}
\sigma_{\alpha}^{CN}(E_{ex},J^\pi)$, can be determined using an appropriate
optical model for the $a$-nucleus interaction. 
 Width fluctuation corrections have been omitted to simplify the notation in
 Equation~\ref{eq:DesReact}, but are included in the calculations.

$G_{\chi}^{CN}(E_{ex},J^\pi)$ is the probability that the CN decays via the
exit channel $\chi$. For reactions that emit one particle (neutron, proton,
alpha, etc.) it depends on the convolution of the transmission coefficient
$T^J_{\chi l_c j_\chi}$ with the level density $\rho_{j_C}(U) $ for the
residual nucleus, divided by analogous terms for all competing decay modes
$\chi'$:
\begin{equation}\label{Eq:DecayBranching} 
 G_{\chi}^{CN}(E_{ex},J^\pi) = 
 \frac{ 
 \sum_{l_c j_\chi j_C} \int T^J_{\chi l_c j_\chi} \rho_{j_C}(U) dE_{\chi}
 } {
    \sum_{\chi' l_c' j_\chi' j_C'} \int T^J_{\chi' l_c' j_\chi'} (E_{\chi'}) 
    \rho_{j_C'} (U') dE_{\chi'}
}. 
\end{equation} 
The quantities $l_c$ and $l_c'$ are the relative orbital angular momenta in the
exit channels. $\vec{j_\chi} = \vec{j_c} + \vec{j_C}$ is the exit channel
spin, related to the total spin $\vec{J} = \vec{l_a} + \vec{j_\alpha} =
\vec{l_c} + \vec{j_\chi}$ by conservation of momentum with the entrance channel
spin, $\vec{j_\alpha} = \vec{j_a}+\vec{j_A}$. $\rho_C(U,j_C)$ is the density
of levels of spin $j_C$ at energy $U$ in the residual nucleus.

Contributions from decays to discrete levels and to regions described by a
level density have to be accounted for and are implicitly included in the
integrals in both the numerator and denominator of Eq.
\eqref{Eq:DecayBranching}. For reactions that involve sequential decays, e.g.
the emission of two neutrons in (n,2n), Eq. \eqref{Eq:DecayBranching} is
repeatedly applied: First, to determine the possible outcomes of the CN decay
in the first step of the emission chain, and second, to follow the subsequent
decays of the intermediate compound nuclei created. In HF calculations, the
final cross sections are obtained by tracking all possible decays in this
manner. All sums over quantum numbers must respect parity conservation, although
this is not explicitly expressed here.

In this paper, we focus on neutron-induced reactions, \idest $\alpha = n+A$.
For such reactions, the optical model potential, used to calculate the first
factor in Eq. \eqref{eq:DesReact}, is well approximated by a one-body
potential~\cite{Gadioli:92}. By far the greatest source of uncertainty comes
from the decay probabilities, a fact that can be attributed to uncertainties in
the nuclear structure inputs. \textit{ab initio} shell-model calculations can
provide nuclear structure information for nuclei with only a dozen or so
nucleons, and traditional shell-model calculations cover a limited number of
nuclei, primarily near closed shells, containing up to around 100 nucleons.
Mean-field and beyond-mean field approaches cover a wider range of nuclei, but
calculating the relevant structure quantities (level densities and gamma-ray
strength functions) is nontrivial. While much progress has been made toward
achieving microscopic nuclear structure inputs for HF calculations of
medium-mass and heavy nuclei, many isotopes needed for applications and for
simulating stellar environments are currently out of reach.

In the absence of microscopic predictions of structural properties,
phenomenological models are used for nuclear level densities and electromagnetic
transition strengths, with parameters that are fitted to available data. Much
effort has been devoted to generate global or regional parameter systematics
\cite{Capote:09} that can be utilized as to perform HF calculations and build
nuclear reaction evaluations~\cite{Koning2005,TENDL:2015,YAHFC:18,Brown:18}.
Alternatively, it is possible to use surrogate reaction data to obtain experimental
constraints on the decay probabilities.

\subsection{Full modeling of the surrogate reaction}
 \label{sec_formalism_full}

In a surrogate experiment, such as the one schematically shown in
Figure~\ref{Fig_n2nSurrogate}, the compound nucleus $B^*$ is produced by an
inelastic scattering or transfer reaction $d+D$ $\rightarrow$ $b$ + $B^*$, and
the desired decay channel is observed in coincidence with the outgoing particle
$b$ at angle $\theta_b$.

The probability for forming $B^*$ in the surrogate reaction (with specific
values for $E_{ex}$, $J$, $\pi$) is $F_{\delta}^{CN}(E_{ex},J,\pi,\theta_b)$,
where $\delta$ refers to the surrogate reaction $d+D$ $\rightarrow$ $b$ + $B^*$.
The quantity 
\begin{equation} 
  P_{\delta\chi}(E_{ex},\theta_b) = \sum_{J,\pi}
  F_{\delta}^{CN}(E_{ex},J^\pi,\theta_b) \;\; G_{\chi}^{CN}(E_{ex},J^\pi) \; ,
\label{Eq:SurReact} 
\end{equation} 
which gives the probability that the CN $B^*$ was formed with energy $E_{ex}$
and decayed into channel $\chi$, can be obtained experimentally by detecting a
discrete $\gamma$-ray transition characteristic of the residual nucleus (or some
other suitable observable).

The distribution $F_{\delta}^{CN}(E_{ex},J,\pi,\theta_b)$, which may be very
different from the CN spin-parity populations following the absorption of a
neutron in the desired reaction, has to be determined theoretically, so that the
branching ratios $G_{\chi}^{CN}(E_{ex},J^\pi)$ can be extracted from the
measurements.

In practice, the decay of the CN is modeled using a Hauser-Feshbach-type decay
model and the $G_{\chi}^{CN}(E_{ex},J^\pi)$ are obtained by adjusting parameters
in the model to reproduce the measured probabilities
$P_{\delta\chi}(E_{ex},\theta_b)$. Subsequently, the sought-after cross section
for the desired (neutron-induced) reaction can be obtained by combining the
calculated cross sections $\sigma_{n+A}^{CN}(E_{ex},J^\pi)$ for the formation of
$B^*$ (from $n$+$A$) with the extracted decay probabilities
$G_{\chi}^{CN}(E_{ex},J^\pi)$, see Eq.~\eqref{eq:DesReact}. Modeling the CN decay
begins with an initial (``prior'') description of structural properties of the
reaction products (level densities, branching ratios, internal conversion
rates), plus a fission model for cases which involve that decay mode. Finally,
a procedure for fitting the parameters of the decay models, e.g. via a Bayesian
approach as introduced in Ref.~\cite{Escher:18prl}, needs to be implemented to
determine the desired cross section, along with uncertainties.

This procedure was recently employed to determine cross sections for neutron
capture on the stable $^{90}$Zr and $^{95}$Mo isotopes (for benchmark purposes),
as well as for neutron capture on the unstable $^{87}$Y nucleus
~\cite{Escher:18prl,Ratkiewicz:19prl}.  It was also used to simultaneously infer
the $(n,\gamma)$ and low-energy (n,f) cross sections for
$^{239}$Pu~\cite{PerezSanchez:20}

Such a full treatment of a surrogate experiment is challenging: It involves
taking into account differences in the angular momentum $J$ and parity $\pi$
distributions between the compound nuclei produced in the desired and Surrogate
reactions, as well as their effect on the decay of the compound nucleus.
Predicting the spin-parity distribution $F_{\delta}^{CN}(E_{ex},J,\pi,\theta_b)$
resulting from a Surrogate reaction is a nontrivial task since a proper
treatment of direct reactions leading to highly excited states in the
intermediate nucleus $B^*$ involves a description of particle transfers, and
inelastic scattering, to unbound states. In addition, a complete treatment
should include consideration of width fluctuation corrections and the possible 
decay prior to reaching equilibrium.

For capture cross sections, it was shown that this type of approach is needed to
account for the spin-parity mismatch in the surrogate
experiment~\cite{EscherDietrich:10,Chiba:10}, while for fission applications it
often suffices to employ the much simpler \we or ratio
approximations~\cite{EscherDietrich:06}.

\subsection{Weisskopf-Ewing approximation for neutron-nucleus reactions and 
surrogate coincidence probabilities}
\label{sec_formalism_we}

The Hauser-Feshbach expression for the cross section of the desired
neutron-induced reaction, Eq. \eqref{eq:DesReact}, conserves total angular
momentum $J$ and parity $\pi$. Under certain conditions the branching ratios
$G_{\chi}^{CN}(E_{ex},J^\pi)$ can be treated as independent of $J$ and $\pi$ and
the cross section for the desired reaction simplifies to 
\begin{eqnarray}
\sigma^{WE}_{n+A,\chi}(E_a) &=& 
\sigma^{CN}_{n+A}(E_{ex}) \;
\mathcal{G}^{CN}_{\chi}(E_{ex}) 
\label{Eq:WELimit} 
\end{eqnarray} 
where $\sigma^{CN}_{n+A}(E_{ex}) =\sum_{J\pi} \sigma^{CN}_{n+A} (E_{ex},J^\pi)$ 
is the cross section describing the formation of the compound nucleus at energy
$E_{ex}$ and $\mathcal G^{CN}_{\chi}(E_{ex})$ denotes the $J\pi$-independent
branching ratio for the exit channel $\chi$. This is the Weisskopf-Ewing limit
of the Hauser-Feshbach theory~\cite{Gadioli:92}. 

The \we limit provides a simple and
powerful approximate way of calculating cross sections for compound-nucleus
reactions. In the context of surrogate reactions, it greatly simplifies the
application of the method. In section \ref{sec_formalism_full} we described the 
process required to obtain the $J\pi$-dependent branching ratios 
${G}^{CN}_{\chi}$ from measurements of $P_{\delta\chi}(E_{ex})$. In the \we
limit, and because $\sum_{J\pi} F_{\delta}^{CN}(E_{ex},J^\pi)=1$,
\begin{equation}\label{Eq:WElimit2}
P_{\delta\chi}(E_{ex}) = {\mathcal G}^{CN}_{\chi}(E_{ex}). 
\end{equation}
Calculating the direct-reaction probabilities
$F_{\delta}^{CN}(E_{ex},J,\pi,\theta_b)$ and modeling the decay of the compound
nucleus are no longer required in this approximation. (In actual applications,
experimental efficiencies have to be included when determining
$P_{\delta\chi}(E_{ex})$; these are omitted for simplicity here, but are
accounted for in the analysis of surrogate experiments.)

The conditions under which the approximate expressions \eqref{Eq:WELimit}) and
\eqref{Eq:WElimit2} are obtained from equations \eqref{eq:DesReact} and
\eqref{Eq:SurReact} are discussed in the appendix.

In addition, the Weisskopf-Ewing approximation can be used in situations in
which the surrogate reaction produces a spin distribution that is very similar
to that of the desired reaction, i.e. 
\begin{align}\label{equalspins}
	F_\delta^{CN}(E_{ex},J^\pi) &\approx F_{n+A}^{CN}(E_{ex},J^\pi),
\end{align}
where,
\begin{align}
 F_{n+A}^{CN}(E_{ex},J^\pi) \equiv
	\frac{ \sigma_{n+A}^{CN}(E_{ex},J^\pi) } {\sum_{J\pi'}
	\sigma_{n+A}^{CN}(E_{ex},J^{\pi'}) },
\end{align} 
since the weighting of the $J^\pi$-dependent decay probabilities in
the measured $P_{\delta\chi}(E_{ex})$ is the same as the weighting relevant to
the desired reaction. While some intuitive arguments have been forwarded in
favor of specific surrogate reaction mechanisms that might satisfy the condition
\eqref{equalspins}, not much is actually known about what spin-parity
distributions $F_\delta^{CN}$ are obtained when producing a CN at high
excitation energies ($E_{ex} >$ 5 MeV) via inelastic scattering or a transfer
reaction. We therefore investigate both the dependence of realistic decay
probabilities $G_{\chi}^{CN}(E_{ex},J^\pi)$ on spin and parity
(Section~\ref{sec_method_strategy}) and the impact of using the \we
approximation in situations in which $G_{\chi}^{CN}(E_{ex},J^\pi)$ depends on
spin and parity (Section~\ref{sec_method_surrogateJPi}).

\section{Assessing the validity of the Weisskopf-Ewing Approximation}
\label{sec_method}

As discussed in the previous section, there are two scenarios in which it is
clearly valid to employ the \we approximation in the analysis of a surrogate
experiment: (a) The decay probabilities $G_{\chi}^{CN}(E_{ex},J^\pi)$ are
independent of \jpi for the decay channel $\chi$ of interest; or (b) The
surrogate and desired reactions produce identical spin distributions
(``serendipitous'' or ``matching'' approach~\cite{Escher:12rmp}). In addition,
there are some intermediate situations in which a \we analysis can give a good
approximation to the true cross section. For instance, it is possible that the
decay probabilities $G_{\chi}^{CN}(E_{ex},J^\pi)$ are only moderately sensitive
to $J\pi$, and that the surrogate and desired reactions populate somewhat
similar \cn spins and parities, so that violations of the \we limit may have
little impact on the extracted cross section. Investigations into the
possibility of using the \we approximation must therefore consider {\em both}
the behavior of the decay probabilities $G_{\chi}^{CN}(E_{ex},J^\pi)$ for the
decay channel $\chi$ of interest {\em and} their influence in typical surrogate
reaction analyses.

Earlier studies, which have done that, demonstrated that it is not {\em a
priori} clear whether the \we limit applies to a particular reaction in a given
energy regime~\cite{EscherDietrich:06,Forssen:07,EscherDietrich:10,Chiba:10}.
For fission applications, it was found that using the \we approximation gives
reasonable cross sections, with violations of the \we limit occurring primarily
at low energies ($E_n$ below 1-2 MeV) and at the onset of first and
second-chance fission~\cite{EscherDietrich:06}. For neutron capture reactions,
however, the $G_{\gamma}^{CN}(E_{ex},J^\pi)$ were found to be very sensitive to
the \jpi and no circumstances have been identified so far in which the \we limit
can be used to obtain capture cross sections~\cite{EscherDietrich:10}.

In the present study we focus on the proposed use of the surrogate method to
determine $(n,n^{\prime})$ and $(n,2n)$ cross sections. To study the validity
of the \we approximation, we proceed in two steps:

\ben \item Investigation of the \jpi dependence of the decay probabilities
$G_{\chi}^{CN}(E_{ex},J^\pi)$ for $\chi = $ $1n$ and $2n$, \idest for one- and
two-neutron emission. \item Assessment of the impact of the \jpi dependence of
the $G_{\chi}^{CN}(E_{ex},J^\pi)$ on cross sections extracted by using
the \we approximation. \een

\subsection{Method for determining spin-parity dependence}
\label{sec_method_strategy}

In the first step, we obtain the $G_{\chi}^{CN}(E_{ex},J^\pi)$ from
well-calibrated Hauser-Feshbach calculations that involve the relevant decay
channels. We selected $n + ^{90}$Zr, $n + ^{157}$Gd, and $n + ^{238}$U as
representative cases for neutron reactions on spherical and deformed nuclei,
with the uranium case representing a nucleus for which fission competes with
particle evaporation and $\gamma$ emission.

For each nucleus, we carried out a full Hauser-Feshbach calculation of the
neutron-induced reaction and calibrated the model parameters to give an overall
good fit of the known neutron cross sections. This local optimization of model
parameters allows us to isolate the spin-parity effects from model
uncertainties. Our optimization procedure accounted for pre-equilibrium effects
using the two-exciton model \cite{KONING200415}, and other competing decay channels.
This is necessary to accurately and realistically reproduce the data without
biasing the model-space parameters. In contrast, the calculations described in
this and the following section include only contributions from \cn decay. This
is consistent with the goal of investigating the ability to determine the
compound cross section from a \we analysis of surrogate data. 

The calculations were carried out with Hauser-Feshbach codes {\sc
Stapre}~\cite{Uhl:76} and {\sc YAHFC-MC}~\cite{YAHFC:18}. The results discussed
here are obtained using the latter. We extracted the branching ratios 
$G_{xn}^{CN}(E,J^\pi)$ for one- and two-neutron emission ($x=1$ and 2, 
respectively) for a range of spin and parity values of the initially formed 
compound nuclei $^{91}$Zr$^*$, $^{158}$Gd$^*$, and $^{238}$U$^*$, and 
investigated their behavior as a function of the excitation energy $E_{ex}$ of 
the CN. Our findings are discussed in Section~\ref{subsec_results_decayProbs}.

\subsection{Method for demonstrating impact of spin-parity dependence}
\label{sec_method_surrogateJPi}

In the second step, we employ the decay probabilities
$G_{xn}^{CN}(E_{ex},J^\pi)$ extracted above to simulate the results of possible
surrogate measurements. This is done by calculating the coincidence
probabilities given by equation \eqref{Eq:SurReact}, which are ordinarily
measured in a surrogate experiment, by multiplying the
$G_{xn}^{CN}(E_{ex},J^\pi)$ with several schematic spin-parity distributions
$F^{CN}_{\delta}(E_{ex},J^\pi)$, summed over all relevant spins and parities:
\bea P_{xn}^{sim}(E_{ex}) &=& \sum_{J\pi}
F^{CN}_{\delta}(E_{ex},J^\pi) G_{xn}^{CN}(E_{ex},J^\pi). \eea 
We normalized the distributions $\sum_{J\pi} F^{CN}_{\delta}(E_{ex},J^\pi) = 1$ 
and did not consider angle dependencies. Multiplication of these simulated 
coincidence probabilities $P_{xn}^{sim}(E_{ex})$ by the CN-formation cross 
section $\sigma^{CN}_{n+A}(E_{ex})$ then yields cross sections
$\sigma^{WE}_{(n,n^{\prime})}(E_n)$ and $\sigma^{WE}_{(n,2n)}(E_n)$ that
correspond to a \we analysis of the simulated surrogate measurement: 
\begin{equation}
   \sigma^{WE}_{(n, xn)}(E_n) = \sigma^{CN}_{n+A}(E_{ex}) P^{sim}_{xn}(E_{ex})
\end{equation} 
for $x=1,2$. In Section~\ref{subsec_results_WE}, we compare the
so extracted cross sections for various spin-parity distributions
$F^{CN}_{\delta}$ to each other and to the known desired cross sections.

To select relevant \jpi distributions for our study, we briefly summarize what
is known about \jpi distributions that typically occur in neutron-induced as
well as surrogate reactions.

\subsubsection{Spin-parity distributions in neutron-induced reactions.}

\begin{figure}[htb] \begin{center}
\includegraphics[width=8.6cm]{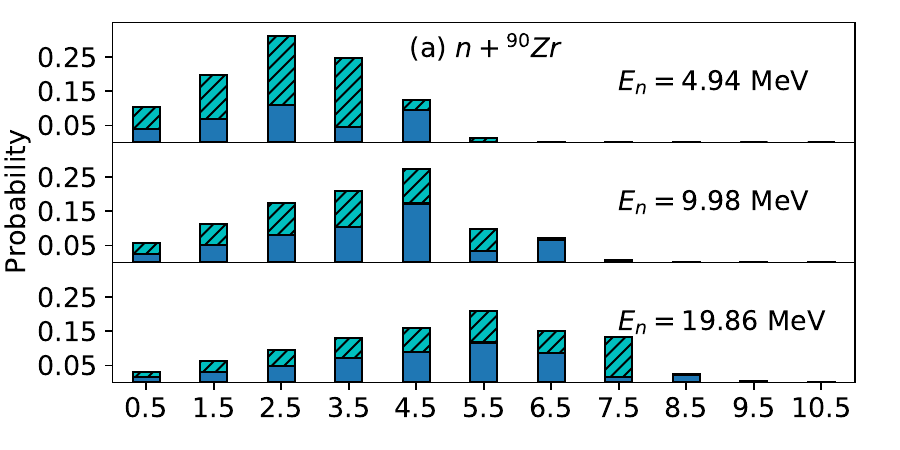}
\includegraphics[width=8.6cm]{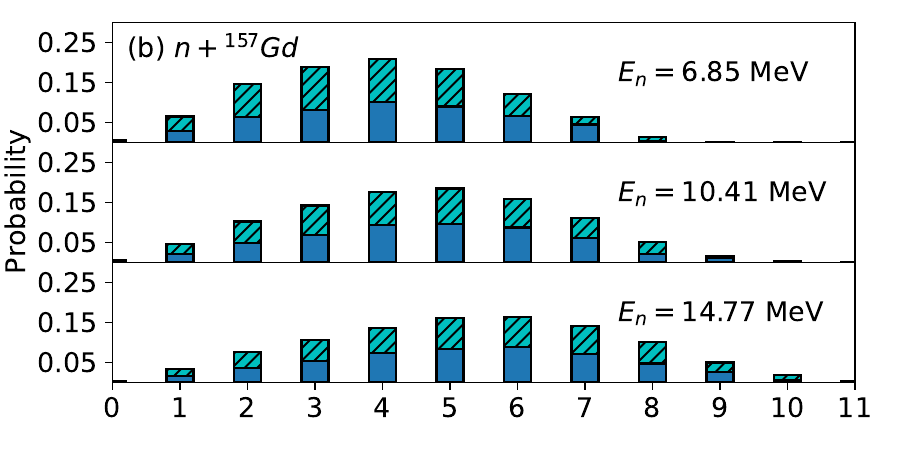}
\includegraphics[width=8.6cm]{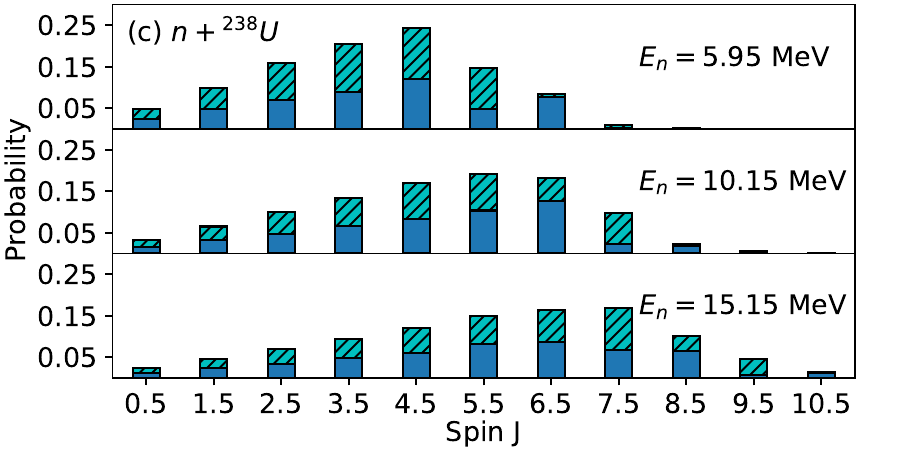}
\end{center} 
\caption{Spin-parity distributions for compound nuclei produced in
neutron-induced reactions, for several neutron energies $E_n$. 
Solid bars are positive- and hatched bars are negative-parity probabilities.
Panels (b) n+$^{157}$Gd and (c) n+$^{238}$U are
representative of deformed rare-earth and actinide nuclei, respectively, while
panel (a) presents the case of a near-closed shell nucleus, n+$^{90}$Zr. Neutron
energies below 1 MeV are important for neutron capture
reactions~\cite{EscherDietrich:10}. For the $(n,n^{\prime})$ and $(n,2n)$
applications considered in this paper, neutron energies between about 5 and 20
MeV are relevant.} \label{Fig_nInducedSpins} \end{figure}

Figure~\ref{Fig_nInducedSpins} shows spin-parity distributions relevant to
neutron-induced reactions, as predicted by calculating the compound-formation
cross sections for various spins and parities, at the energies indicated. For
Zr, a spherical optical-model calculation is sufficient, while rare earths and
actinides require coupled-channels treatments, which can be carried out by
suitably deforming a spherical optical model, see 
Ref.~\cite{Nobre:2014,Nobre:2015},
or by using a coupled-channels scheme that is specifically adjusted for the
nucleus or nuclear region of interest, see Refs.~\cite{Soukhovitski:2016,
Soukhovitski:2020,MASLOV200477,Dietrich:12,EscherDietrich:10}. We have used the
Koning-Delaroche optical model \cite{Koning:03} for Zr and Gd, and Soukhovitskii
~\cite{Soukhovitski:2016,Soukhovitski:2020} for the U.

For the $(n,n^{\prime})$ and $(n,2n)$ applications
considered here, neutron energies between about 5 and 20 MeV are relevant. The
examples selected here involve target nuclei with low spins ($3/2^-$ for
$^{157}$Gd and $0^+$ for the even-even $^{90}$Zr and $^{238}$U nuclei), so the
spin-distributions are closely connected to the angular-momentum transferred in
the reaction.

Panel (a) shows the population of positive and negative parity states for the
n+$^{90}$Zr example, for several neutron energies $E_n$. At $E_n \approx$ 1 MeV,
p-wave capture dominates~\cite{Forssen:07} and produces a distribution that
favors negative-parity states within a narrow range of spins. As the energy
increases, contributions from higher partial waves result in smoother
distributions, centered at larger angular momentum values, and with a more equal
partition between positive and negative spins.

Panels (b), for n+$^{157}$Gd, and (c), for n+$^{238}$U, are representative of the
situations one encounters for deformed rare-earth and actinide nuclei,
respectively. Overall, the distributions are smoother for the deformed nuclei
than for the Zr case and involve larger values of angular momentum. With
increasing $E_n$, the positive and negative parity distributions become similar,
while at low energies, $E_n < $ 1 MeV, the distributions can look quite
different from each other~\cite{EscherDietrich:10}.

\subsubsection{Spin-parity distributions in surrogate reactions.}

The findings of the following illustrate that it is {\em not} correct to assume
that the spin-parity distribution of a \cn produced in a surrogate reaction is
given by the spin and parity behavior of the level density for that nucleus.
The reaction mechanism plays a critical role in selecting which states act as
doorways into the compound nucleus. The population of these doorway states
determines the \jpi distribution for the surrogate reaction.

Figure~\ref{Fig_n2nSurrogate} illustrates schematically the excitation energies
that a surrogate reaction has to populate in order to produce decay information
relevant to $(n,\gamma)$, $(n,n^{\prime})$ and $(n,2n)$ reactions. For neutron
capture, $E_{ex}$ values between about 5 and 10 MeV have to be reached, for
inelastic scattering, energies between approximately 10 and 20 MeV are relevant,
and for $(n,2n)$ reactions, $E_{ex} =$ 20-30 MeV are important. These energy
regimes exhibit high level densities, and transfer reactions aiming to populate
these energy ranges are very different from those used for traditional nuclear
structure studies. It should therefore not surprise that standard DWBA or even
coupled-channels calculations cannot be used to reliably calculate the direct
(surrogate) reactions that produce such states.

\begin{figure}[htb] \begin{center}
\includegraphics[width=8.6cm]{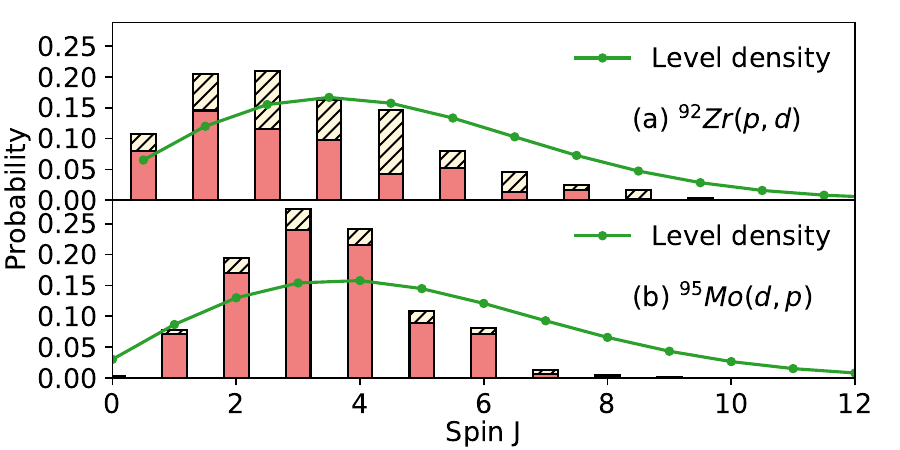}
\end{center} 
\caption{
Spin-parity distributions (bars) near the neutron separation energy, as
predicted for use with specific surrogate experiments. Solid bars are
positive-parity and hatched bars are negative-parity probabilities. Panel (a)
shows the half-integer J distribution in the compound nucleus $^{91}$Zr$^*$
resulting from a $^{92}$Zr(p,d) reaction with $E_p$ = 28.5 MeV at $E_{ex}$ =
7.25 MeV \cite{Escher:18prl}. Panel (b) shows the integer valued result for
$^{95}$Mo$(d,p)$ surrogate reaction with $E_d$ = 12.4 MeV at $E_{ex}$ = 9.18
MeV \cite{Ratkiewicz:19prl}. In both cases, the predicted spin-parity
distributions were used in combination with models for the decay of the
respective compound nuclei, leading to the successful determination of
(benchmark) neutron capture cross sections. For comparison, the spin
distribution calculated from an energy-dependent level density model, which
assumes equal parity distribution, is given by the green solid curve
\cite{vonEgidy:09}.} 
\label{Fig_SurrSpins} 
\end{figure}

Predicting the spin-parity distributions for these higher excitation energies
requires taking into account both the surrogate reaction mechanism and the
nuclear structure at these higher energies. For instance, to calculate the \jpi
population in the \cn $^{91}$Zr$^*$ that was produced via the $^{92}$Zr(p,d)
pickup reaction in a recent surrogate experiment with $E_p$ = 28.5
MeV~\cite{Escher:18prl}, it was necessary to consider the structure of deep
neutron hole states, which exhibit considerable spreading \cite{Duhamel:91, 
Escher:12rmp}. Furthermore,
two-step mechanisms involving $(p,d^{\prime})(d^{\prime},d)$ and
$(p,p^{\prime})(p^{\prime},d)$ combinations of inelastic scattering and pickup
contribute significantly to the reaction. These have a strong influence on the
final spin-parity distribution in $^{91}$Zr$^*$~\cite{Escher:18prl}, which is
shown for $E_{ex}$ = 7.25 MeV in Figure~\ref{Fig_SurrSpins}(a).
The influence of the reaction mechanism is reflected in the differences 
between the predicted spin-parity population (bars) and the spin 
distribution in a representative level density model at the same 
excitation energy (green curve). 

Around the neutron separation energy, \idest in the energy region of interest to
neutron capture, the angular behavior of the (p,d) cross section was found to be
fairly structureless, and the \jpi distribution was seen to vary little over
several MeV around $E_{ex}$ = $S_n(^{91}$Zr$)=$ 7.195 MeV~\cite{Escher:18b}.
These observations reflect the fact that the surrogate reaction does not produce
a simple single-particle excitation, but populates specific {\em doorway} states
which mix with neighboring complex many-body states to form the compound
nucleus.

The $(d,p)$ transfer reaction, which -- at first glance -- seems to be a
well-matched surrogate for neutron-induced reactions, turns out to involve
non-trivial reaction mechanisms as well. The case of interest is that in which
the deuteron breaks up in the combined Coulomb-plus-nuclear field, and the
neutron is absorbed while the proton escapes and is observed in a
charged-particle detector. Calculating the resulting \cn \jpi distribution
requires a theoretical description that separates elastic from nonelastic
breakup and, in principle, one also needs to separate out inelastic breakup,
rearrangement, and absorption. This challenge has generated strong interest in
developing a more detailed formalism for inclusive $(d,p)$
reactions~\cite{Lei:15a, Lei:15b, Potel:15, Carlson:16, Potel:17}. This
formalism was used to calculate the \jpi distribution relevant to the
$^{95}$Mo$(d,p)$ surrogate reaction described in Ref.~\cite{Ratkiewicz:19prl}.
The calculated \jpi distribution, for excitation energies near the neutron
separation energy in $^{95}$Mo is shown in Figure~\ref{Fig_SurrSpins}(b).
Here, again, the predicted spin-parity distribution (bars) does not follow the 
distribution of spins that are expected to be available at this energy, based 
on a representative level density model (green curve).

Inelastic scattering with charged light ions is a third type of reaction that
has been employed in surrogate reaction
measurements~\cite{Scielzo:10a,Ressler:11a,Hughes:14,Ota:15b,PerezSanchez:20}.
From these experiments, as well as from traditional studies of giant
resonances~\cite{Bertrand:80,Bonetti:84,Martin:79}, it is known that inelastic
scattering can produce a compound nucleus at a wide range of excitation
energies. There is evidence that this type of reaction is also likely to
produce \jpi distributions that are broad and may be centered at angular
momentum values of 5-10 $\hbar$ \cite{Scielzo:10a,PerezSanchez:20}.
Furthermore, for inelastic $\alpha$ scattering, a staggering of even and odd
parity populations is expected, since the reaction populates predominantly
natural-parity states.

\subsubsection{Schematic spin-parity distributions}
\label{sec_method_schematicSpins}

\begin{figure}[htb] \begin{center}
\includegraphics[width=8.6cm]{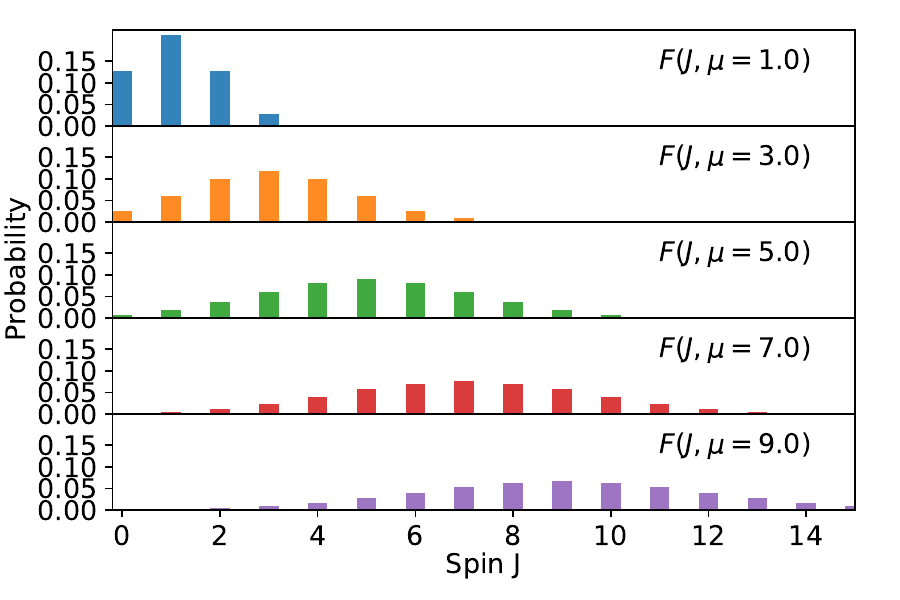}
\end{center} 
\caption{
 Schematic spin distributions
employed in the current study. Each is of the form $F(J,\mu) \propto {\cal N}
(m=\mu, sd=\sqrt{\mu})$, where $\cal N$ is a normal distribution and mean spin
$\mu$ is indicated in the legend. The spin values $J$ are either integer or
half-integer, for even-$A$ and odd-$A$ nuclei, respectively, and equal probability
is assigned to positive and negative parity states.}
\label{Fig_SchematicSurrSpins} \end{figure}

In order to investigate the impact of a spin-parity mismatch between the
desired and surrogate reaction on the cross section obtained from a \we
analysis, we employ the schematic distributions $F_{\delta}^{CN}(J^\pi)$ shown
in Figure~\ref{Fig_SchematicSurrSpins}. We include distributions that are
centered at both low and high angular-momentum values and allow for more
spread-out distributions in the latter case. The distributions centered at low
$J$ values allow us to investigate situations in which the surrogate reaction
populates lower spins than the desired reaction. Variations in parity are not
explicitly considered for this part of the sensitivity study, as we found the
decay probabilities to be less sensitive to parity than to variations in spin.

The distributions shown will be combined with the decay probabilities
$G_{\chi}^{CN}(E_{ex},J^\pi)$ extracted from our calibrated Hauser-Feshbach
calculations (see Section~\ref{subsec_results_decayProbs}) to simulate a range
of possible surrogate data $P_{\delta\chi}(E_{ex},\theta)$ using
Eq.~\eqref{Eq:SurReact}. For simplicity, we will neglect the energy dependence
of the \jpi distributions. This should be a reasonable approach for our
sensitivity studies, as recent results indicate that these distributions vary
slowly with energy \cite{Ratkiewicz:19prl,Escher:18prl}.

\section{Results} \label{sec_results}

We first demonstrate that the one- and two-neutron decay probabilities depend on
the spin, and to a lesser extent, the parity of the compound nucleus. The
dependence is strongest at low energies and for spherical nuclei, and lesser at
higher energies and for deformed nuclei. Then, we show the impact of the \we
approximation on the outcome of simulated surrogate experiments, giving insight
into the effect that the spin dependence has on predicted cross sections.

\subsection{Decay probabilities for representative nuclei}
\label{subsec_results_decayProbs}

\begin{figure}[htb]
\includegraphics[width=8.6cm]{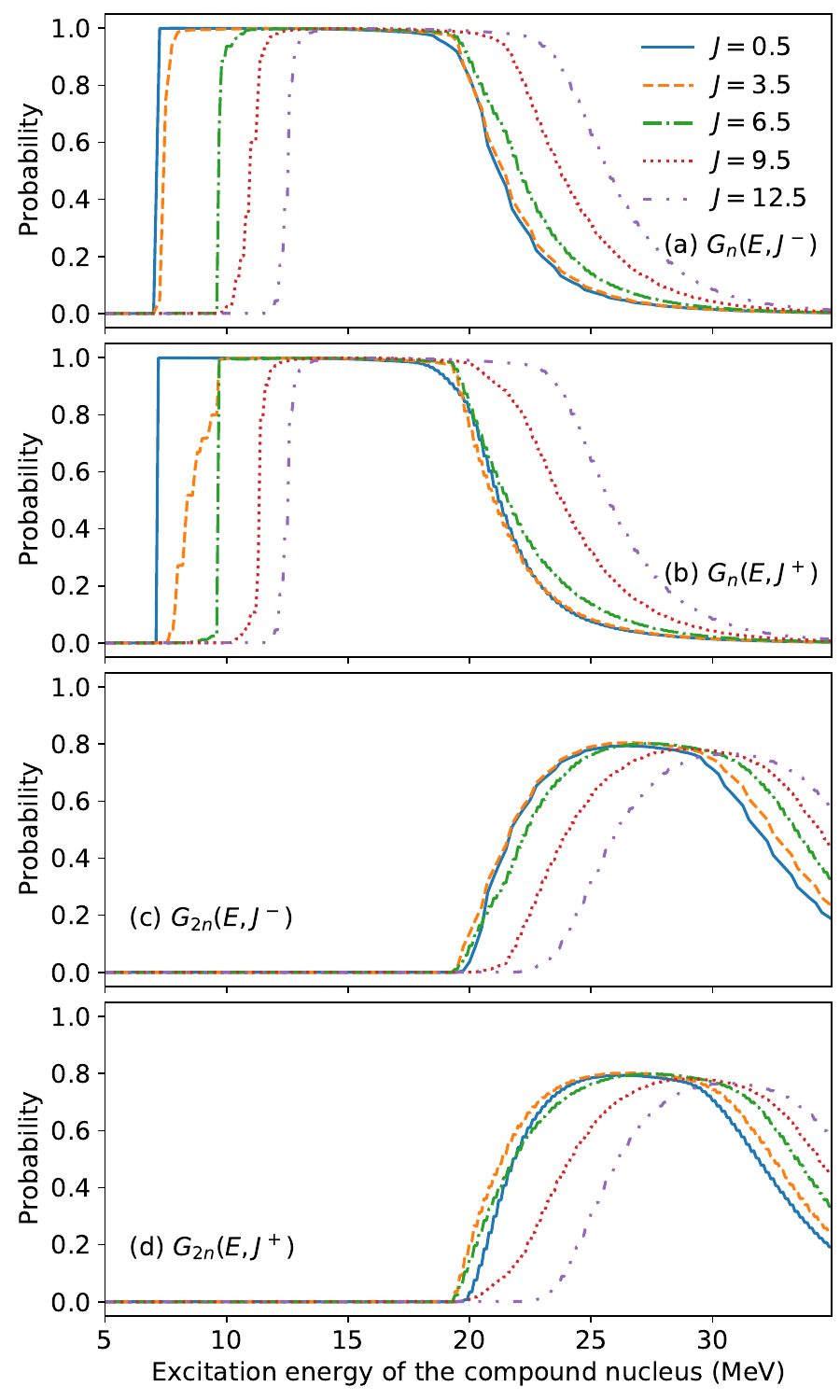}
\caption{
Probabilities for neutron emission from the
$^{91}$Zr$^*$ nucleus, as function of excitation energy, for various \jpi values
of the compound nucleus. Both decay channels exhibit a strong dependence on the
spin of the compound nucleus. The variance is seen to be greatest at the onset
of one-neutron emission, near $E_{ex}=S_n(^{91}$Zr$)=7.194$ MeV. } 
\label{Fig_decayProbs_zirconium}
\end{figure}

$G_{xn}^{CN}(E_{ex},J^\pi)$ for one- and two-neutron emission from the compound
nucleus $^{91}$Zr$^*$ are shown in Figure~\ref{Fig_decayProbs_zirconium}, for
both positive and negative parities and a variety of spins. The behavior of
$G_{xn}^{CN}(E_{ex},J^\pi)$ just above the CN separation energy, corresponding
to $E_{ex}=S_n(^{91}$Zr$)$ $=7.194$ MeV, is governed by the interplay of the
neutron-transmission coefficients and the low-energy structure of the residual
nucleus $^{90}$Zr which is reached by one-neutron emission. The situation is
schematically illustrated in Figure~\ref{Fig_n2nSurrogate}. Due to the shell
structure of the nucleus, the low-energy spectrum of $^{90}$Zr is very sparse,
with the first excited state occurring at 1.76 MeV. Since both the ground
state and the first excited state have $J^{\pi} = 0^+$ and s- and p-wave
neutron emission dominates at low energies, the residual nucleus can only be
reached from low-spin states in the compound nucleus $^{91}$Zr$^*$. This
suppression of neutron emission from all but the lowest spin states in
$^{91}$Zr$^*$ is well known from earlier studies of neutron capture reactions,
and a dependence on parity is observed as well
~\cite{Forssen:07,Escher:12rmp,Boutoux:13}.

As the excitation energy in $^{91}$Zr$^*$ increases, additional states in the
residual nucleus become accessible and the decay probabilities
$G_{xn}^{CN}(E_{ex},J^\pi)$ for higher $J$ values take on non-zero values. In
the region between $E_{ex} = 15 - 20$ MeV, the one-neutron emission probability
is essentially unity, because of the weakness of competing decay channels.

In the energy region between 20 and 27 MeV, we observe the transition from
predominantly one-neutron emission to two-neutron emission. We see significant
dependence of the branching ratio on the spins of the compound nucleus for $J\ge
6.5$, while there is much weaker dependence for $J\le 6.5$. The decay
probabilities are not very sensitive to parity. 

\begin{figure}[htb] 
\includegraphics[width=8.6cm]{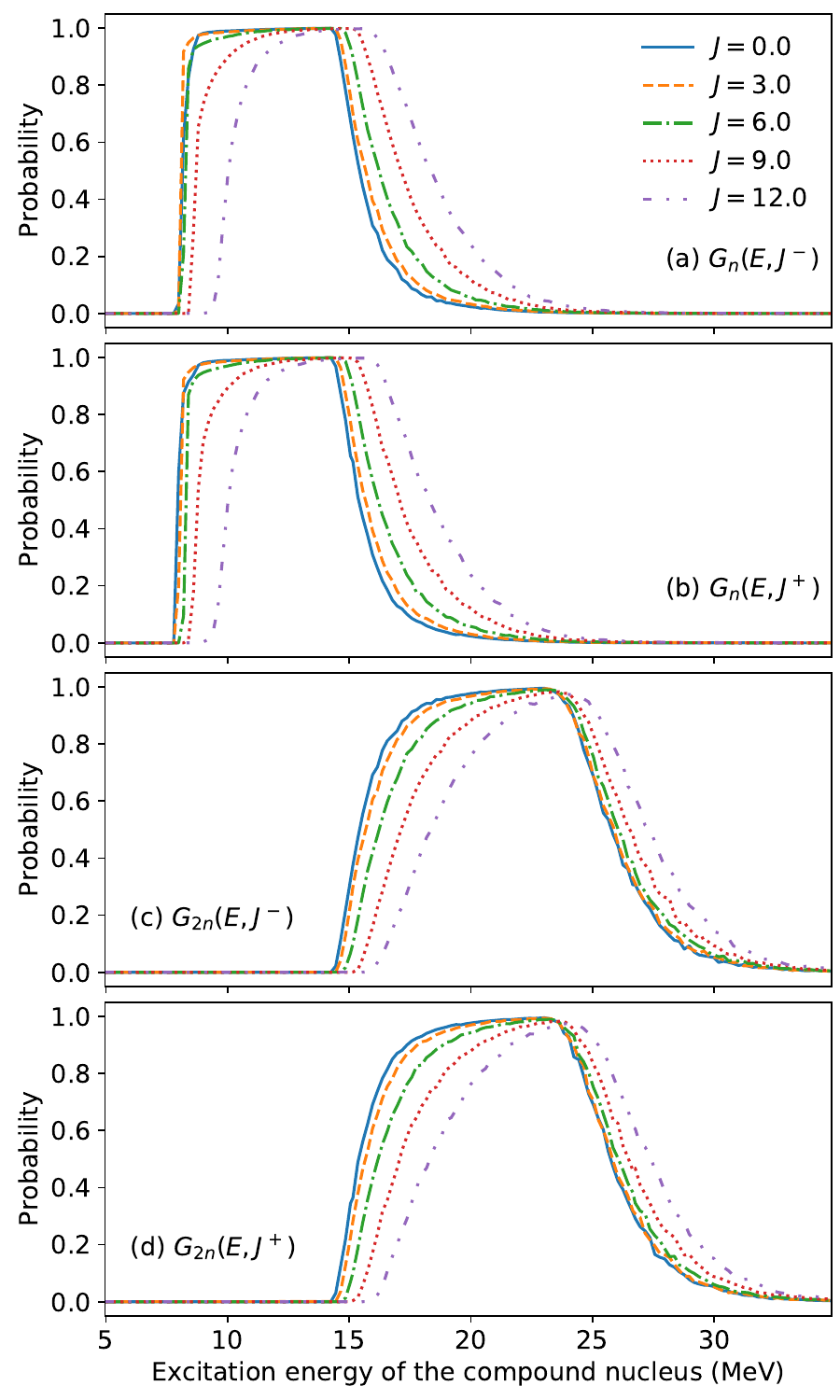}
\caption{
Probabilities for one-and two-neutron emission from the
$^{158}$Gd$^*$ nucleus, as function of excitation energy, for various \jpi
values of the compound nucleus. The decay probabilities for both channels are
seen to depend on the angular-momentum states populated in the compound nucleus,
at the onset of one-neutron emission near $E_{ex}=S_n(^{158}$Gd$)=7.937$ MeV 
and in the transition region where the
two-neutron channel opens. } \label{Fig_decayProbs_gadolinium} \end{figure}

\begin{figure}[htb] \begin{center}
\includegraphics[width=8.6cm]{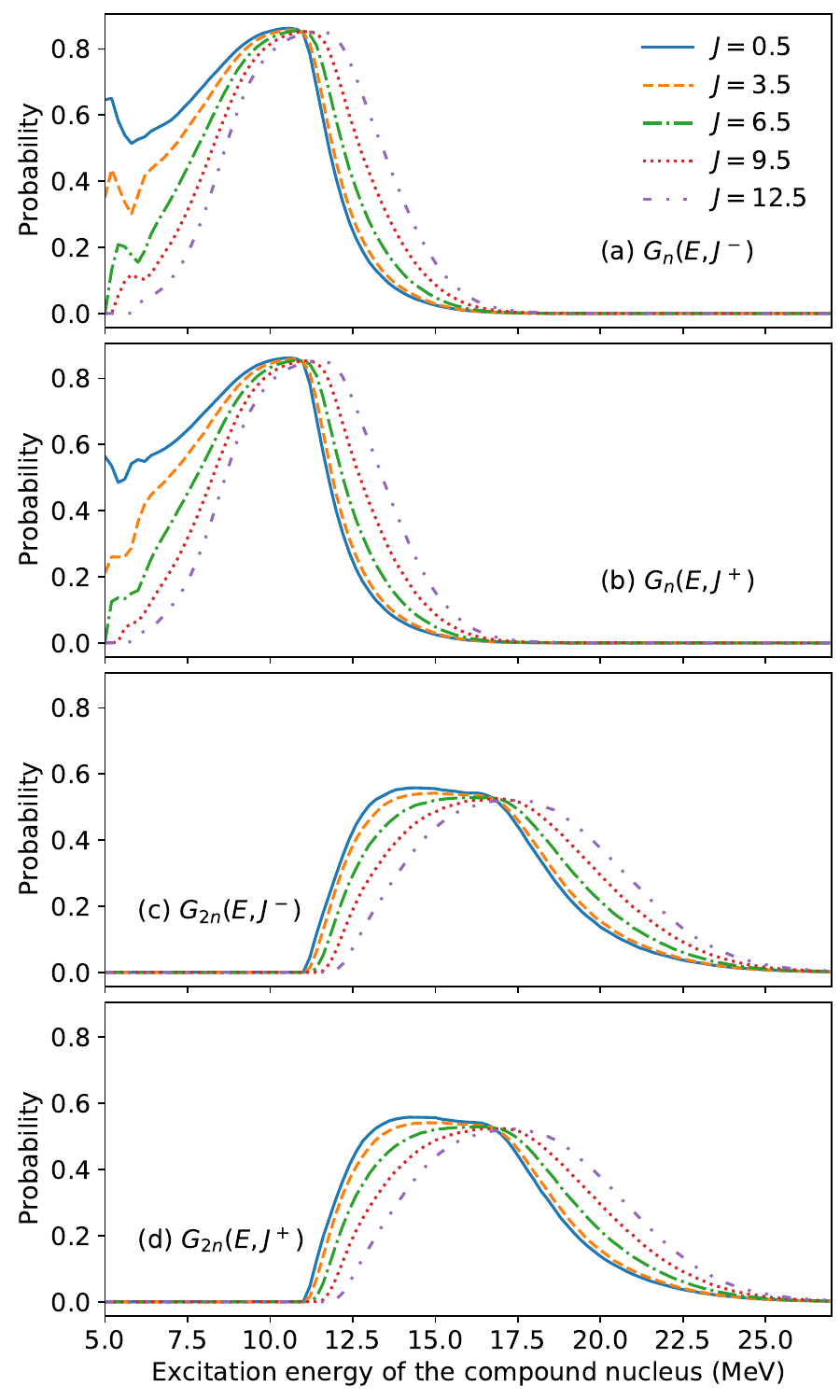}
\end{center} 
\caption{
Probabilities for one-and two-neutron emission from the
$^{239}$U$^*$ nucleus, as function of excitation energy, for various \jpi values
of the compound nucleus. We observe a strong spin- and parity-dependence of
$G_{1n}^{CN}(E_{ex},J^\pi)$ near $E_{ex}=$ $S_n(^{239}$U$)$ = 4.806 MeV, which
lies just below the threshold for fission. } \label{Fig_decayProbs_uranium}
\end{figure}

Figure~\ref{Fig_decayProbs_gadolinium} shows the analogous one- and two-neutron
emission probabilities for the decay of the rare-earth nucleus $^{158}$Gd.
Here, the dependence on spin is weaker than in the Zr case, especially near the
one-neutron separation energy of the compound nucleus. This is primarily due to
the significantly higher level density in the gadolinium nuclei: While the first
excited state in $^{90}$Zr is at 1.76 MeV, there are 15 levels below 0.5 MeV in
$^{157}$Gd. In general, the level densities in deformed nuclei are much higher,
and the sensitivity of the \cn decays to spin and parity is reduced. This is
also true at higher energies: The competition between one- and two-neutron
emission shows significant dependence on the compound-nuclear spins, although
the sensitivity is not as strong as in the zirconium case. 
Figure~\ref{Fig_decayProbs_uranium} shows the one-
and two-neutron emission probabilities for the $^{239}$U nucleus. Like the
gadolinium case discussed, the uranium nuclei are deformed and have a much
higher level density than the zirconium nuclei: $^{238}$U has 16 levels below 1
MeV. The transition from one-neutron to two-neutron emission, which lies near
the threshold for second-chance fission, is also sensitive to the angular
momentum population of the compound nucleus. Multiple channels compete at all
energies considered and no clear plateaus for the probabilities emerge, unlike
in the other cases considered. 

For all three cases discussed, we have observed that there is enhanced
sensitivity of the neutron emission probabilities near the thresholds. It can
therefore be expected that a failure to account for the spin-parity mismatch in
the analysis of surrogate reaction will result in extracted $(n,n^{\prime})$ and
$(n,2n)$ cross sections that do not reflect the true threshold behavior. This
will be investigated in more detail in the next subsection.

\subsection{Impact of spin dependence of 1n and 2n decay probabilities} 
\label{subsec_results_WE}

In the previous section, we observed that the one- and two-neutron decay
probabilities show a significant dependence on the spin of the compound nucleus
and a lesser dependence on parity. Here we study the impact of this dependence
on cross sections obtained under the assumption of the validity of the \we
approximation. We use the schematic spin distributions
$F_{\delta}^{CN}(E_{ex},J^\pi)$ discussed in
Section~\ref{sec_method_schematicSpins}. They are conveniently parameterized as
discretized normal distributions with mean $\mu$ and variance $\sigma^2 = \mu$: \bea
F_{\delta}^{CN}(E_{ex},J^\pi) \propto {\cal N} (m=\mu, sd = \sqrt{\mu}).\eea 
The distributions are cut off above $J=50$ and normalized to unity. For
the even-even compound nucleus $^{158}$Gd$^*$, we consider the five
distributions, $\mu = 1, 3, 5, 7, 9$, shown in
Figure~\ref{Fig_SchematicSurrSpins}; for the odd nuclei $^{91}$Zr$^*$ and
$^{239}$U$^*$ we use $\mu = 1.5, 3.5, 5.5, 7.5$, and $9.5$.

\begin{figure}[htb] \begin{center}
\includegraphics[width=8.6cm]{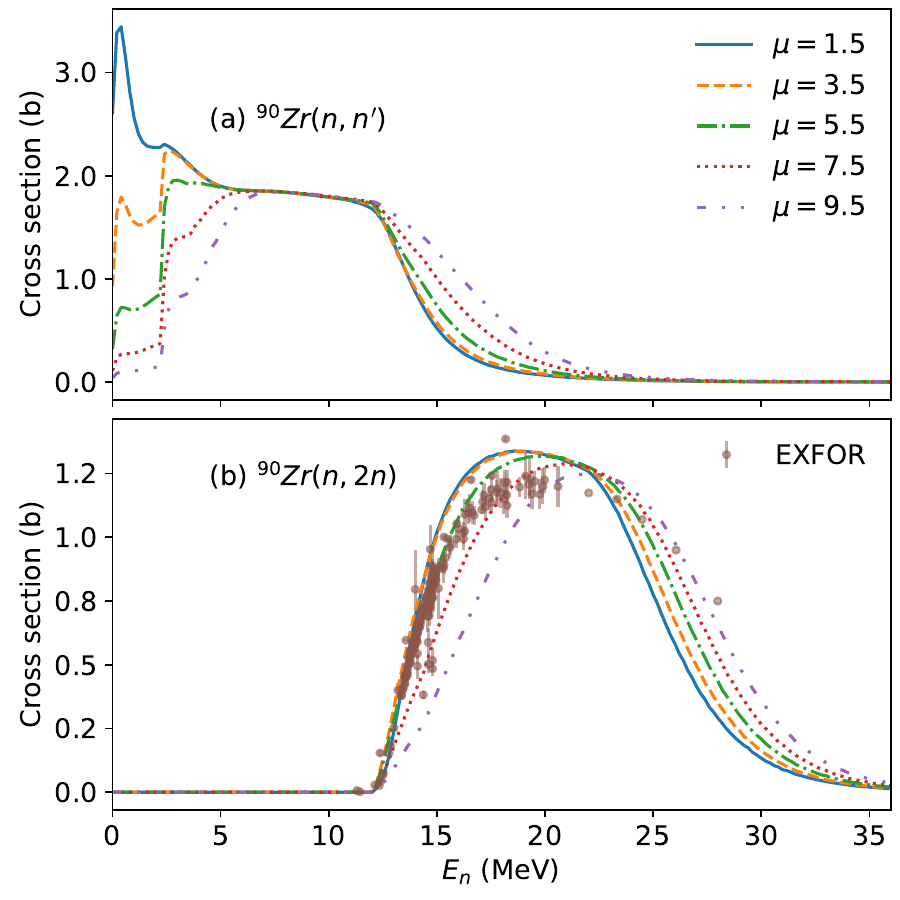}
\end{center} 
\caption{
Cross sections for (a) $^{90}$Zr$(n,n^{\prime})$ and (b) $^{90}$Zr$(n,2n)$,
obtained from simulated surrogate data, using the \we assumption. The
underlying schematic spin-parity distributions used are indicated in the legend.
The shape of the transition depends clearly on which simulated surrogate data is
used, with the cross sections varying by $\pm 30\%$ at about $E_{n} = 15$ MeV.
The $^{90}$Zr$(n,2n)$ cross section varies by $\pm 4\%$ near its maximum, which
is located at about $E_{n} = 20$ MeV. For comparison, experimental
data~\cite{EXFOR:18} for $^{90}$Zr$(n,2n)$ is shown in panel (b). The only 
data for the inelastic scattering case is for scattering to an isomeric state. }
\label{Fig_WE_zirconium} \end{figure}

Results for $^{90}$Zr$(n,n^{\prime})$ and $^{90}$Zr$(n,2n)$ cross sections
obtained from a \we analysis of the simulated surrogate data are shown in
Figure~\ref{Fig_WE_zirconium}. As expected, the threshold regions for both
reactions are particularly sensitive to spin effects. At the onset of inelastic
scattering, it is not possible to obtain a reliable $(n,n^{\prime})$ cross
section; both shape and magnitude show a very large variance. Different spin
distributions give the same magnitude of this cross section in the region of the
plateau, but there is again significant uncertainty in the region where the
two-neutron channel opens up.

\begin{figure}[htb] \begin{center}
\includegraphics[width=8.6cm]{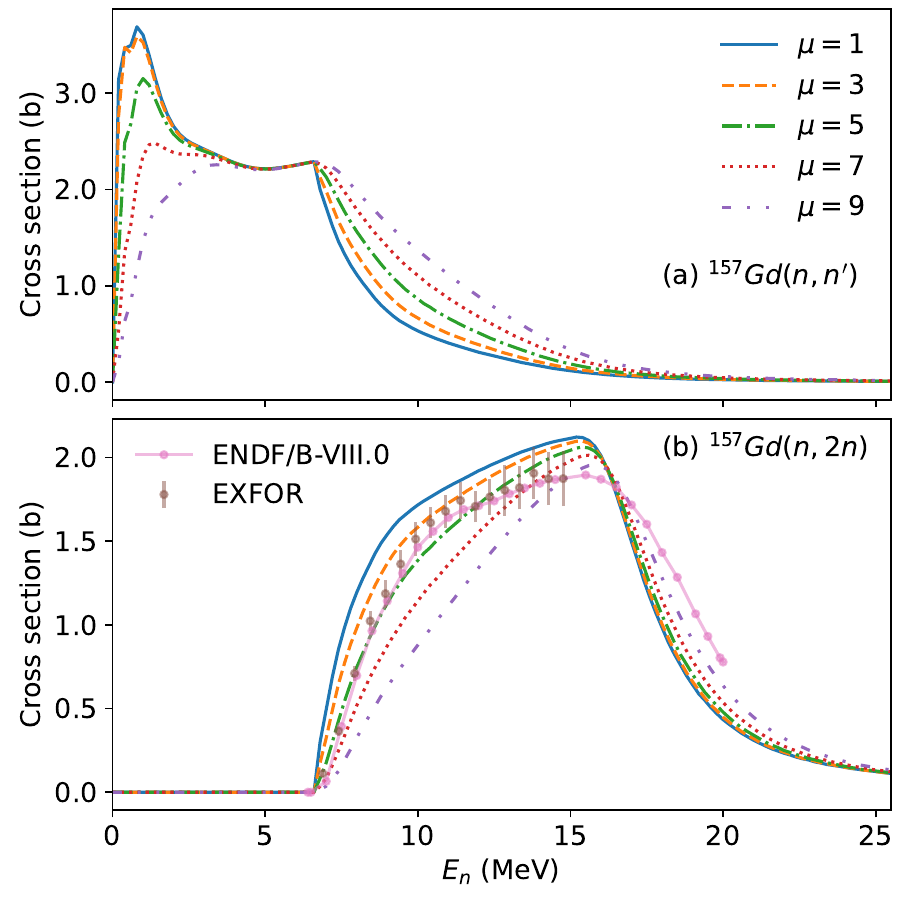}
\end{center} 
\caption{
Cross sections for $^{157}$Gd$(n,n^{\prime})$ and $^{157}$Gd$(n,2n)$, obtained
from simulated surrogate data, using the \we assumption and several schematic
spin-parity distributions. In the energy region where the transition from one-
to two-neutron emission occurs, the cross sections exhibit greater uncertainty,
varying by $\pm 57\%$ for $(n,n')$ and $\pm 13\%$ for $(n,2n)$ at $E_n = 10$
MeV. The maximum for $(n,2n)$ near $E_n=15$ MeV, the variation is $\pm 62\%$
for $(n,n')$ and $\pm 1\%$ for $(n,2n)$. For comparison, directly measured data
~\cite{EXFOR:18} is shown for the $^{157}$Gd$(n,2n)$ cross section; no data are
available for the inelastic cross section. } \label{Fig_WE_gadolinium}
\end{figure}

Given the findings in the previous section, we expect the situation to be better
for the gadolinium case, shown in Figure~\ref{Fig_WE_gadolinium}. While the
$^{157}$Gd$(n,n^{\prime})$ cross section near the onset of inelastic scattering
varies less than the analogous zirconium cross section, it is still quite
unreliable. The value of the $^{157}$Gd$(n,n^{\prime})$ cross section shows no
dependence on the simulated spin-parity distribution in a region around $E_n=5$ 
MeV. Not surprisingly, the \we approximation for different sets of simulated
surrogate data yields results that are consistent with each other in an energy
regime where there is little to no competition from other decay channels. The
maximum for the $^{157}$Gd$(n,2n)$ cross section occurs near $E_n=15$ MeV, where
the different sets of surrogate data differ from each other by about $4 \%$,
which is an uncertainty that is similar to the error bands obtained from direct
measurements. Overall, it appears that the \we approximation might provide a
very rough estimate of the $(n,2n)$ cross section of a rare earth nucleus.

\begin{figure}[htb] \begin{center}
\includegraphics[width=8.6cm]{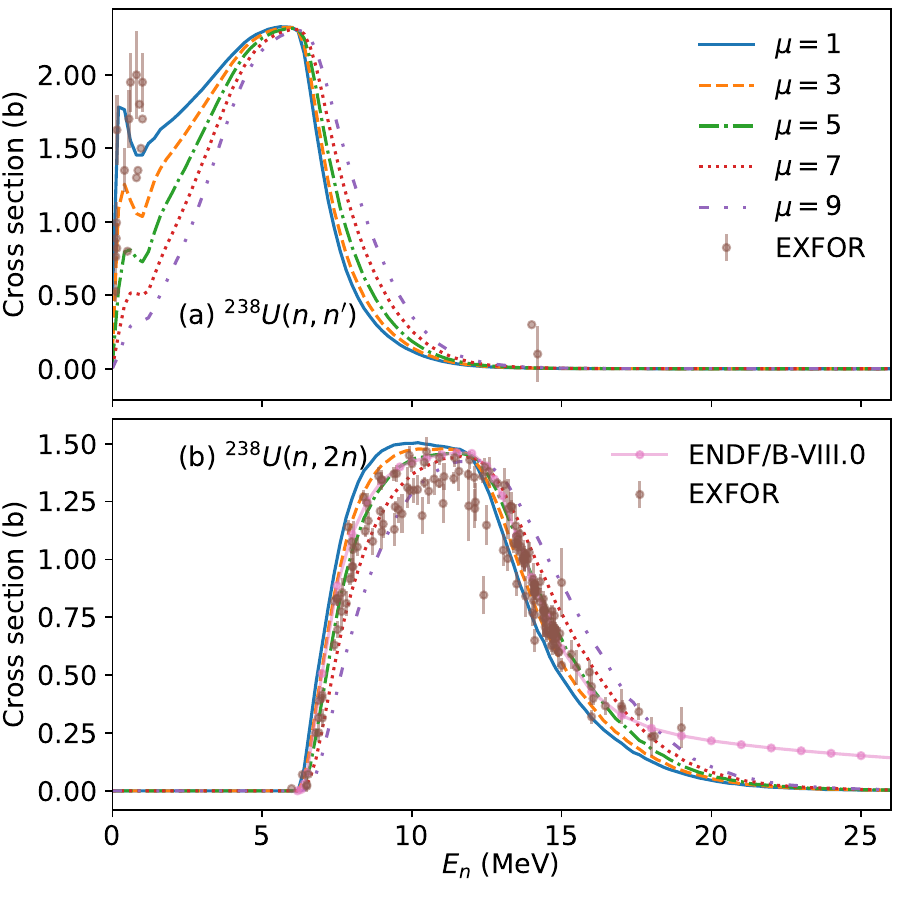}
\end{center} 
\caption{
Cross sections for (a) $^{238}$U$(n,n^{\prime})$ and (b) $^{238}$U$(n,2n)$,
obtained from simulated surrogate data, using the \we assumption and several
schematic spin-parity distributions. The $^{238}$U$(n,2n)$ results agree
reasonably well with the existing data~\cite{EXFOR:18}. For the inelastic case,
data is only available data for low energies, where direct reaction mechanisms
are known to contribute. } \label{Fig_WE_uranium} \end{figure}

For the uranium case, shown in Figure~\ref{Fig_WE_uranium}, we observe a further
decrease in sensitivity to differences in spin. Even so, the shape of the
$^{238}$U$(n,n^{\prime})$ cross section cannot be reliably extracted at low
energies. With increasing energy, the \we approximation appears to become more
reliable.  In fact, the $^{238}$U$(n,2n)$ cross section obtained from the
simulated data are in good agreement with available directly-measured data. At
energies above 18 MeV, however, where no data exists, we see deviations from the
ENDF evaluation\cite{Brown:18}. We attribute this to the neglect of
pre-equilibrium contributions, which are included in evaluations but neglected
in standard WE analysis. 

Overall, we find that the \we approximation can provide rough first estimates
for the $(n,2n)$ cross sections of nuclei with large level densities, such as
rare earth and actinide nuclei, while the low-energy behavior is much less
reliable. Specifically, near thresholds there is clearly increased sensitivity
of the decay to the underlying spin-parity distribution in the compound nucleus.
As a result, the shape of the extracted cross sections do not reproduce the true
cross sections very well. Notably, the \we approximation fails at the onset of
one-neutron emission. This is in line with earlier findings about the
limitations of this approximation for neutron capture cross sections.

In addition, it should be stressed that we have focused on the compound
contributions to the $(n,n^{\prime})$ and $(n,2n)$ cross sections here. For
inelastic scattering, it is well known that direct (pre-equilibrium) mechanisms
provide significant additional contributions, which are not considered here.
These contributions are known to affect the spins populated in the target
nucleus \cite{PhysRevC.75.054612, PhysRevC.104.044605} and will exacerbate the
deficiencies of the WE approximation.
These have to be calculated separately and added to the cross section, similar
to what is done for the direct-reaction component in an evaluation.
Unfortunately, for many nuclei there is little data available for neutron
inelastic scattering, and the calculations are challenging, so this reaction
channel requires additional studies, both experimentally and theoretically.

\section{Outlook} \label{sec_outlook}

We have investigated the potential use of the \we approximation for determining
$(n,n^{\prime})$ and $(n,2n)$ cross sections from surrogate reaction data.
Earlier work for neutron-induced fission and radiative neutron capture
demonstrated that this approximation yields reasonable approximations for the
fission cross sections, but fails for capture, making it necessary to employ
more detailed theoretical modeling in the latter case.

We modeled the nuclear structure properties that determine the decay of a \cn
via $1n$ and $2n$ emission, as well as the combined effect of the nuclear
structure and the surrogate reaction mechanisms on the cross-section results
that one obtains from a \we analysis of the indirect data. We found that the
Weisskopf-Ewing approximation fails to give consistent cross section shapes in
the presence of a spin-parity mismatch between the desired and surrogate
reactions. The outcomes are worse for nuclei with low level density, \idest for
lighter nuclei and for those in regions near closed shells. While rough
estimates for the cross sections might be obtained for $(n,2n)$ reactions on
well-deformed rare-earth and actinide nuclei, we find that nuclei in the mass-90
region are more sensitive to the effects of spin and parity. Furthermore,
inelastic neutron scattering cross sections are found to be quite sensitive to
angular-momentum effects and thus require a detailed treatment of the reaction
mechanism, similar to that recently used for extracting capture cross sections
from surrogate data.

Suggestions to find a surrogate reaction that approximates the spin-parity
distribution relevant to the desired reaction are well-motivated, as the use of
the \we approximation greatly simplifies surrogate applications. However, not
enough is known about the angular momentum and parity of the compound states
that are populated in a surrogate reaction to plan an appropriate experiment.
Recent work has demonstrated that the surrogate reactions that produce a
compound nucleus at the high energies of interest involve higher-order reaction
mechanisms, which render inadequate the type of simple angular-momentum estimates that are
often used in traditional nuclear structure studies. It is also not
necessarily true that a surrogate reaction produces spins in a compound nucleus
that are higher than those relevant to neutron-induced reactions. This means
that in order to achieve cross section results with appropriate shapes and
errors less than about 30\%, surrogate reaction data will need to be combined
with full modeling of the reaction mechanism, as described in section
\ref{sec_formalism_full}.

In light of our findings that the \we approximation is insufficient for
determining $(n,n')$ and $(n,2n)$ cross sections, we believe that further
development of surrogate reaction theory is important for addressing existing
nuclear data needs. Inelastic scattering $(n,n')$ reactions in particular are
poorly constrained by direct measurement techniques. Alternative indirect
methods~\cite{Larsen:11} do not address $(n,n')$ and $(n,2n)$ reactions. Recent
surrogate reaction applications to neutron capture have demonstrated how to
proceed to accurately extract cross sections from surrogate data in situations
where the \we approximation fails
\cite{Escher:18prl,Ratkiewicz:19prl,PerezSanchez:20}. Given the limited utility
of the \we approximation for neutron induced one- and two- neutron emission
reactions, we conclude that additional developments are needed in order to
describe the relevant reaction mechanisms, such as those involved in the
($^{3}$He,$^{3}$He$^{\prime}$) scattering experiment described in Figure
\ref{Fig_n2nSurrogate}.

\section{Acknowledgements} 

This work was performed under the auspices of the U.S.
Department of Energy by Lawrence Livermore National
Laboratory under Contract No. DE-AC52-07NA27344 with
support from LDRD Project No. 19-ERD-017, and the 
Defense Science and Technology Internship (DSTI) and Glenn
T. Seaborg Institute (GTSI) summer student programs. A part
of this work was supported by DOE Grant No. DE-FG02-
03ER41272.

\appendix
\section{Conditions of the Weisskopf-Ewing limit} \label{sec_appendix}

As discussed in Section~\ref{sec_formalism}, if the decay probabilities
$G_{\chi}^{CN}(E_{ex},J^\pi)$ are independent of spin and parity, or the
surrogate reaction produces a \cn spin distribution which is very similar to
that produced by the neutron-induced reaction, the cross section for the desired
reaction can be obtained very simply as: 
\begin{align} 
 \sigma_{n+A,\chi}(E_{n})
= \sigma_{n+A}^{CN}
	(E_{ex}) P_{\delta\chi}^{CN} (E_{ex}),
\end{align} 
where $P_{\delta\chi}^{CN} (E_{ex})$ is the coincidence probability
determined from the surrogate measurement.

The latter of these options, the `serendipitous' or `matching' condition
requires that $F_\delta^{CN}(J^\pi) \approx$ $F_{n+A}^{CN}(E_{ex},J^\pi)$ holds.
A comparison of $F_{n+A}^{CN}(E_{ex},J^\pi)$ for representative nuclei and
energies $E_{ex}$, shown in Figure~\ref{Fig_nInducedSpins} of this paper and in
Figure 3 of Ref.~\cite{EscherDietrich:10}, with realistic surrogate spin-parity
distributions, such as those shown in Figure~\ref{Fig_SurrSpins}, indicates that
it is difficult to identify and carry out a surrogate reaction experiment that
can achieve this condition.

Here, we briefly review the conditions in which the decay probabilities become
approximately independent of $J^\pi$, \idest $G_{\chi}^{CN}(E_{ex},J^\pi)$
$\rightarrow {\cal G}_{\chi}^{CN}(E_{ex})$ (see also
Refs.~\cite{Gadioli:92,EscherDietrich:06}):

First, the energy of the compound nucleus has to be sufficiently high, so that
almost all channels into which the nucleus can decay are dominated by integrals
over the level density. In that case, the denominator in Eq.
\eqref{Eq:DecayBranching} does not include decays to discrete levels.

Second, correlations between the incident and outgoing reaction channels, which
can be formally accounted for by including width fluctuation
corrections~\cite{Hilaire:03}, have to be negligible. These correlations enhance
elastic scattering, at the expense of the inelastic and reaction cross sections,
and are most prominent at the low energies relevant to capture reactions. Width
fluctuations are negligible if the first condition (above) is satisfied.

Third, the transmission coefficients $T^{J}_{\chi' l'_c j'_\chi}$ associated
with the available exit channels have to be independent of the spin of the
states reached in these channels. This condition is sufficiently well satisfied
since the dependence of transmission coefficients on target spin is very weak
and, in fact, is ignored in many Hauser-Feshbach codes.

Fourth, the level densities $\rho_{j_C}$ in the available channels have to be
independent of parity and their dependence on the spin of the relevant nuclei
has to be of the form $\rho_{j_C} \propto (2{j_C} + 1)$. While level densities
are known to depend on parity, that dependence becomes weaker with increasing
excitation energy and is often ignored in statistical reaction calculations. In
addition, many successful applications use level densities that are parametrized
in a form that is factorized (for each parity) as: 
\begin{align} 
 \rho_{j_C}(U_C) = w(U_C)\frac{(2j_C+1)}{2\sigma_C^2}
  \exp\left(\frac{-j_C(j_C+1)}{2\sigma_C^2}\right), 
\end{align}
where $w(U_C)$ contains the energy dependence of the level density and
$\sigma_C$ is the spin cut-off factor. At low energies ($E_{ex} \leq 3$ MeV),
typical values for $\sigma_C^2$ are 7-10 in the Zr region and 12-16 in the Gd
region~\cite{vonEgidy:08}. As $E_{ex}$ increases from a few MeV to about 20
MeV, $\sigma_C^2$ can increase by a factor 4 or more for these mass
regions~\cite{vonEgidy:09}. If we then assume that the spins populated in the
residual nucleus are small compared to the $\sigma_C$, the level density can be
written as 
\begin{align} 
  \rho_{j_C}(U_C) \approx \frac{w_{C}(U_C)}{2\sigma_C^2}(2j_C+1).
\end{align}

When the above conditions are satisfied, the decay probabilities from Eq.
\eqref{Eq:DecayBranching} take the form: 
\begin{flalign}\label{Gapprox}
&G_{\chi}^{CN}(E_{ex},J^\pi) =  & \nn \\ &\frac{
  \sum_{l_c j_\chi j_C} \int T^J_{\chi l_c j_\chi} w_{C}(U_C)(2j_C+1) dE_{\chi}
} {
 \sum_{\chi' l_c' j_\chi' j_C'} \int
      T^J_{\chi' l_c' j_\chi'} (E_{\chi'}) w_{C'}(U_C')(2j_C'+1) dE_{\chi'}
}. & 
\end{flalign}
We can carry out the sum over $j_C$ if we use the triangle rule
$|j_{\chi}-j_c|<j_C<|j_{\chi}+j_c|$ to obtain the identity 
\begin{align}
	\sum_{j_{C}}(2j_C+1) = (2j_{\chi}+1)(2j_c+1). \nn
\end{align} 
and analogously for the $j_\chi$: 
\begin{align}
	\sum_{j_{\chi}}(2j_\chi+1) = (2J+1)(2l_c+1), \nn
\end{align} 
to obtain the spin-independent decay probabilities:
\begin{flalign}\label{Gapprox2} 
&\mathcal{G}_{\chi}^{CN}(E_{ex}) =   &\\ &\frac{
  \left( \sum_{l_c} (2l_c+1) T_{\chi l_c } \right) \int (2j_c+1) w_{C}(U_C)
  dE_{\chi}
} {
 \left(\sum_{\chi' l_c' }
      (2l_c+1)T_{\chi' l_c'} (E_{\chi'}) \right) \int (2j_c'+1)w_{C'}(U_C')
      dE_{\chi'}
}. & 
\end{flalign}

In summary, in order for the $G_{\chi}^{CN}(E_{ex},J^\pi)$ to become
independent of spin and parity, the energy $E_{ex}$ of the \cn must be high
enough so that decays to the continuum of residual nuclei dominate, and the
reaction must populate spins that are small relative to the spin cutoff
parameter. Since neutron-induced reactions and surrogate reactions can produce
different spin distributions, it is possible that the conditions for the
validity of the \we approximation are satisfied for one type of reaction, but
not the other.

\bibliography{wepaper}

\end{document}